\documentclass[epj]{svjour}
\usepackage{graphics,psfig}
\newcommand{\alfs}{\mbox{$\alpha_s$}}
\newcommand{\mztwo}{\mbox{$M_Z^2$}}
\def\np#1#2#3   {{ Nucl. Phys.} {\bf#1}, #2 (#3). }
\def\pcps#1#2#3 {{ Proc. Cam. Phil. Soc.} {\bf#1}, #2 (#3). }
\def\pl#1#2#3   {{ Phys. Lett.} {\bf#1}, #2 (#3). }
\def\plc#1#2#3   {{ Phys. Lett.} {\bf#1}, #2 (#3); }
\def\prep#1#2#3 {{ Phys. Rep.} {\bf#1}, #2 (#3). }
\def\prev#1#2#3 {{ Phys. Rev.} {\bf#1}, #2 (#3). }
\def\prl#1#2#3  {{ Phys. Rev. Lett.} {\bf#1}, #2 (#3). }
\def\prs#1#2#3  {{ Proc. Roy. Soc.} {\bf#1}, #2 (#3). }
\def\ptp#1#2#3  {{ Prog. Th. Phys.} {\bf#1}, #2 (#3). }
\def\rmp#1#2#3  {{ Rev. Mod. Phys.} {\bf#1}, #2 (#3). }
\def\rpp#1#2#3  {{ Rep. Prog. Phys.} {\bf#1}, #2 (#3). }
\def\zp#1#2#3   {{ Z. Phys.} {\bf#1}, #2 (#3). }
\def\epj#1#2#3   {{ Eur. Phys. Jour.} {\bf#1}, #2 (#3). }

\begin{document}

\title{Studies of Higher Twist and Higher Order Effects in NLO and
 NNLO QCD Analysis of Lepton-Nucleon Scattering Data on $F_2$ and
 $R ={\sigma_L}/{\sigma_T}$}
\author{ U.~K.~Yang and A.~Bodek}

\institute{
Department of Physics and astronomy,
University of Rochester, Rochester, NY 14627 }
\date{Received: date / Revised version: date}
\abstract{
We report on the extraction of the higher twist contributions
to $F_2$ and  $R ={\sigma_L}/{\sigma_T}$ from  global NLO and NNLO QCD fits
to lepton nucleon scattering
data over a wide range of $Q^2$. The NLO fits require both
target mass and higher twist contributions at low $Q^2$. 
However, in the NNLO analysis,
the data are  described by the NNLO QCD predictions
(with target mass corrections) without the need for any significant
contributions from higher twist effects. An estimate of the
difference between NLO and NNLO parton distribution functions
is obtained. }

\maketitle

\section{Introduction}

Within the theory of Quantum Chromodynamics (QCD)
Parton Distribution Functions (PDFs) are extracted
from global fits to deep-inelastic (DIS) structure functions
measured in lepton-nucleon scattering
experiments. At present, PDFs are available from
 QCD fits~\cite{MRSR2,CTEQ4M,GRV94} (to various processes including
the structure
functions $F_2$ and $xF_3$)  performed
to order  $\alfs^2$ (Next-to-Leading Order, NLO)
 in perturbation theory~\cite{NLO}. If the range
of the fits is extended to low momentum transfer ($Q^2$), the data
indicate~\cite{dupaper}
 that both target mass (TM) corrections~\cite{GPtm}
 and higher twist (HT) effects~\cite{renormalon}
must be included.  Understanding of the various contributions
to the nucleon structure functions is important in view of
the high precision data at low $Q^2$ that is about to
come out from HERMES, H1 and Zeus, and from
experiments at the CEBAF facility at Jefferson Lab.
In this communication we show that if the fits
are performed with the inclusion of the higher order~\cite{NNLO}
$\alfs^3$ terms (Next-to-Next-to-Leading, NNLO),
the extracted higher twist terms are small. The higher twist
terms extracted in NLO analyses appear to originate from
missing NNLO terms.

In all QCD calculations, physical observables for
interactions with hadrons
are calculated in terms of integrals of products of hard scattering
coefficient functions and parton distribution functions. 
The coefficient function is a generalization of the Born elastic parton
scattering cross section. All NLO QCD analyses of structure function
use one-loop coefficient functions 
and two loop splitting functions 
(needed for the evolution of PDFs from low to high $Q^2$).
%Many such physical observables
%have been calculated to higher orders (beyond NLO) in QCD. At low values of
%$Q^2$, QCD calculations with higher order corrections
%are available for observables
%such as the Gross-Llewellyn Smith (GLS)
%and Bjorken Sum Rules. 
However, recent precise measurements of DIS sum rules, and
of cross sections
for the production of top quarks,  $W$ and $Z$ bosons
in high energy proton-antiproton 
collisions pose a new challenge to theory and
require better predictions from QCD.
%QCD prediction beyond NLO are available for
%these processes.
The calculations
for these processes require both two-loop coefficient functions
(which have been calculated) and NNLO PDFs.
Unfortunately, NNLO PDFs are not yet available,
because only the first few moments of the three-loop splitting functions have
been calculated to date.
Therefore, it has been the practice to use NLO PDFs as input
to the NNLO expressions.
In this communication we also obtain an estimate of the difference between
NLO and NNLO PDFs in a phenomenological way.
This difference may be used in estimating the
uncertainty in the higher order NNLO QCD calculations from
this source.

\section{NLO QCD analysis}

First, we perform an extraction of the
higher twist contributions within a NLO analysis~\cite{dupaper}.
All DIS $F_2$ proton and deuteron data (SLAC, BCDMS, and NMC)
~\cite{SLACF2,BCDMSF2,NMCF2} are used in this analysis.
We also include very high $x$ SLAC proton data
(between $0.8$ and $0.90$)~\cite{SLACres}
that were not included in any previous analysis.
Corrections for nuclear binding effects~\cite{GOMEZ} in the deuteron are
applied as described in our previous communication~\cite{dupaper}.
We use the MRS(R2) PDF (with a modified d quark distribution~\cite{dupaper}
that fits the large $x$ deuteron data which are corrected for the nuclear
binding effects). We use the Georgi-Politzer calculation~\cite{GPtm}
for the target mass corrections. These involve
using the scaling variable 
$\xi=2x/(1+\sqrt{1+4M^2x^2/Q^2})$ instead of $x$.
Higher twist effects originate when scattering from
a single quark cannot be resolved at lower $Q^2$.
Both the target mass and the higher
twist  effects are suppressed by powers of $1/Q^2$

We extract the magnitude of the higher twist terms
from the data within the framework of the
 renormalon model~\cite{renormalon}.
In the renormalon model approach, the 
model predicts the complete $x$ dependence of the higher twist
contributions 
 to  $F_2$, $2xF_1$ (and therefore  $R ={\sigma_L}/{\sigma_T}$),
 and $xF_3$, with only two unknown parameters
$A_2$ and $A_4$ which
determine the overall level of the  $1/Q^2$
and $1/Q^4$ terms. 
We extract the $A_2$ and $A_4$ parameters,  by fitting to the global
data set for $F_2$ and $R [= F_2(1+4Mx^2/Q^2)/2xF_1 - 1]$.
The values of $A_2$ and $A_4$ for the proton and deuteron
are the same in this model.
The $x$ dependence of non-single part of $2xF_1$ differs from that of $F_2$,
 but is the same as that
of $xF_3$ within $1/Q^2$ power correction.

\begin{figure}
\centerline{\psfig{figure=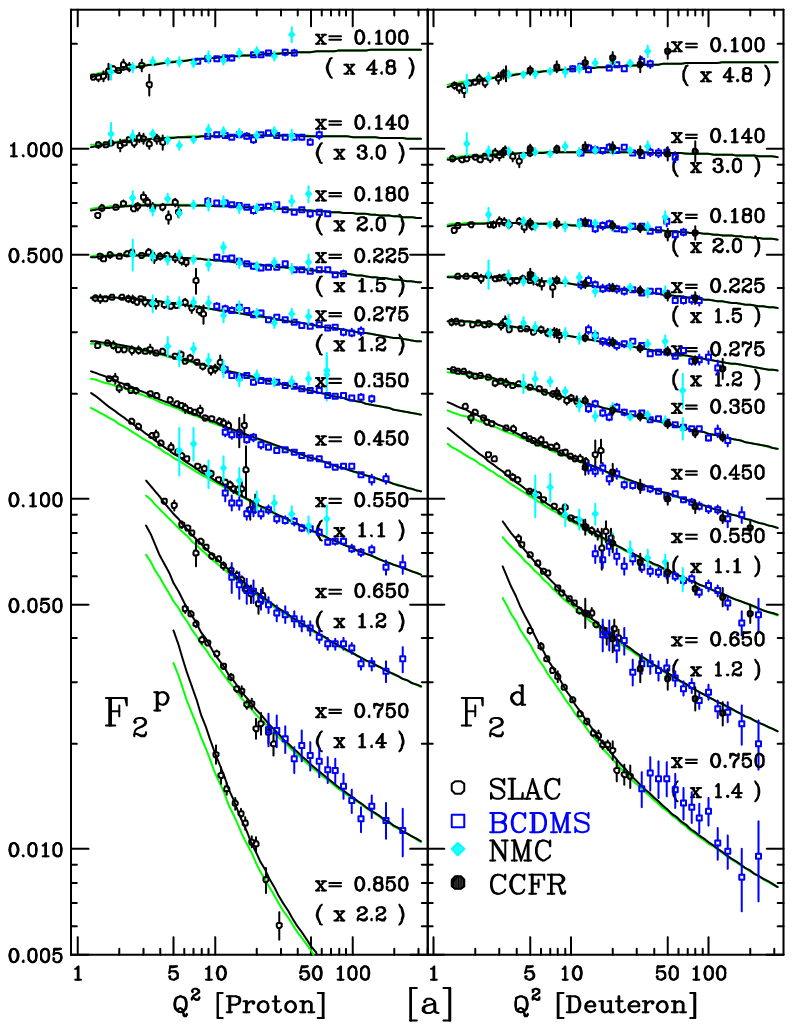,width=3.1in,height=4.1in}}
\centerline{\psfig{figure=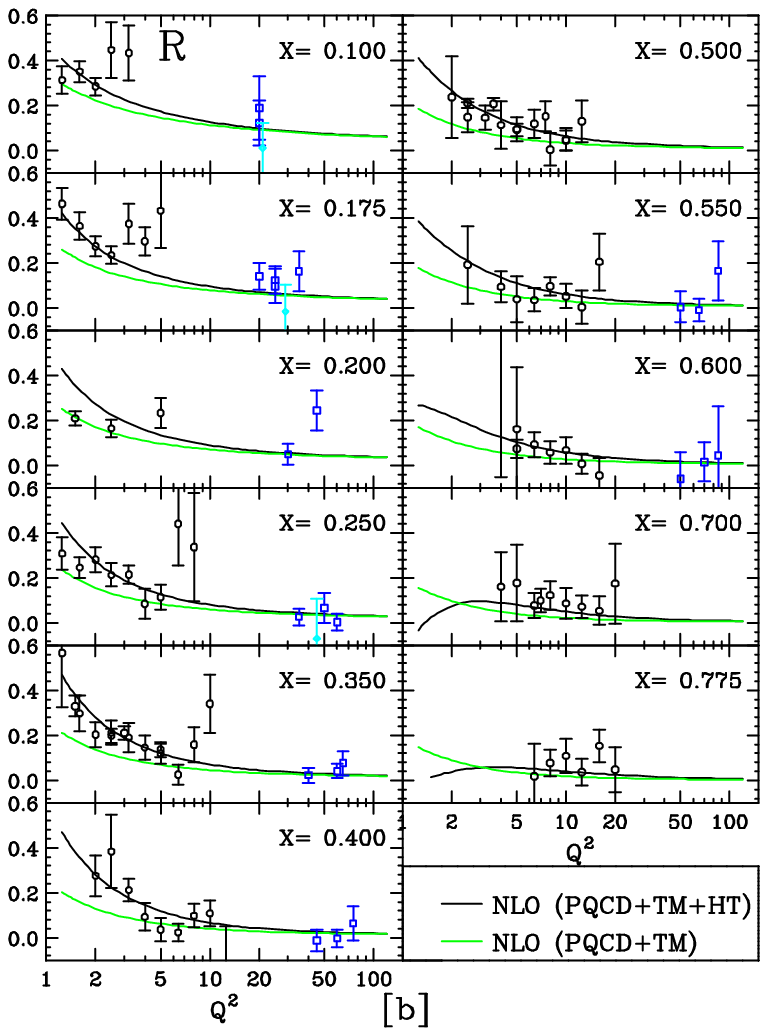,width=3.1in,height=4.1in}}
\caption{Comparison of data with the QCD NLO+TM+HT fit
(the renormalon HT model, $f^{NLO}(x)$, 
and the modified NLO MRS(R2) PDFs are used).
 The CCFR neutrino data are also shown 
for comparison.
(a) Comparison of $F_2$ data and the NLO+TM prediction
with and without HT contributions. 
(b) Comparison of $R$ data and the NLO+TM prediction
with and without HT contributions.}
\label{fig:disht}
\end{figure}

\begin{figure}
\centerline{\psfig{figure=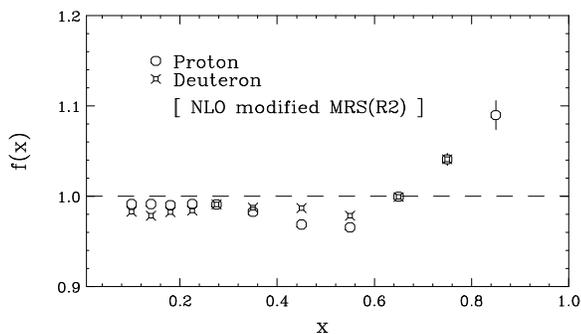,width=3.0in,height=1.7in}}
\caption{The floating factor $f^{NLO}(x)$ as a function of $x$ 
extracted from the NLO analysis with the modified MRS(R2) PDFs. 
This factor is expected to be close
to 1.0 if the modified MRS(R2) PDFs provide a good representation of
the data within the NLO+TM+HT analysis.}
\label{fig:floatNLO}
\end{figure}
The higher
twist coefficients are extracted from a  global fit to all 
DIS $F_2$  in the kinematic region 
($0.1<x<0.90$, $1.25<Q^2<260$ GeV$^2$)
with the following form; 
$F_2 = F_2^{pQCD+TM}[1+ht(x,Q^2)] f^{NLO}(x)$. 
Here $f^{NLO}(x)$ is a floating factor to investigate possible
$x$ dependent corrections to the PDFs. If the PDFs provide
a good representation of  the data, 
%(as they should)
$f^{NLO}(x)$  should be close to 1.0.
We allow a different floating factor for the proton
and deuteron data. The floating factors for the proton and deuteron data
are  an indication on how well the standard PDFs describe the
distribution of $u$ quarks and $d$ quarks, respectively.
The normalization of  BCDMS~\cite{BCDMSF2} and  NMC~\cite{NMCF2} relative 
the SLAC~\cite{SLACF2} data are allowed to float within the errors.
In the case of the BCDMS data, a systematic error shift~\cite{Virchaux}
 $\lambda$ (in standard deviation units) is allowed
to account for the correlated point-to-point  systematic errors.
The higher twist fits with the modified NLO MRS(R2) pQCD
prediction including TM effects
are performed simultaneously
on the proton and deuteron $F_2$ data with 7 free parameters
(two relative normalizations per target, the two higher
twist parameters
and the BCDMS $\lambda$). In addition, 10 floating factors
for the deuteron (at $x$ values of 0.10, 0.14, 0.18, 0.225,
0.275, 0.35, 0.45, 0.55, 0.65, and 0.75)
 and 11 floating factors for the proton
(with the addition of $x$=0.85) are allowed to
vary.

\begin{figure}
\centerline{\psfig{figure=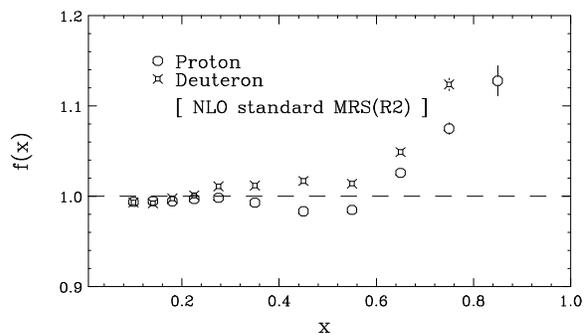,width=3.0in,height=1.7in}}
\caption{The floating factor $f^{NLO}(x)$ as a function of $x$ 
extracted from the NLO analysis with the standard MRS(R2) PDFs.
The larger floating factors for the deuteron than for the proton
indicate that the $d$ quark distribution at high $x$ is underestimated 
in the standard MRS(R2) PDFs.}
\label{fig:floatNLO_nodu}
\end{figure}

 Figure~\ref{fig:disht} shows that the QCD NLO fit with TM effects and
with the
renormalon model higher twists contributions yields a reasonable
description ($\chi^2/d.o.f.=1369/926$, $\chi^2=1245(F_2)+124(R)$ )
 of the $x$ and $Q^2$ dependence for $F_2$ and $R$
with just the two free higher
twist parameters.
The CCFR neutrino data~\cite{CCFRF2}
are shown for comparison though they are not used in the fit.
 The extracted values of $A_2$ and $A_4$
are $-0.100 \pm 0.005$
 and $-0.0024 \pm 0.0007$, respectively.
The contribution of $A_4$ is found  to be negligible in this
NLO analysis. The relative normalization of the NMC and BCDMS data
to the SLAC data, and major systematic error shift 
of the BCDMS data are shown in Table~\ref{ht-norm}.
Figure~\ref{fig:floatNLO}
shows the extracted floating factor $f^{NLO}(x)$ as a function of $x$ for
the proton and deuteron respectively. 
The fact that the
extracted values
are close to 1.0 indicates that the modified NLO MRS(R2) PDFs
provide a good description of the quark distributions. 
%of the $d$ and $u$ quark distributions.
At the highest value of $x$ ($x$=0.85) the floating factor
is higher. Note that because there are no data at very high $Q^2$ at this
value of $x$, there may be a strong correlation between the floating
factor and the higher twist contribution. It is also possible that
the renormalon higher twist model does not account for all the 
physics beyond NLO in the very high $x$ region. 
Since our modifed MRS(R2) PDFs have an enhanced $d$ quark distribution
at high $x$ than the standard MRS(R2) PDFs,
we also investigate the sensitivity of the extracted
 higher twist terms to the choice of the high $x$ $d$ quark
distribution.
We obtain very similar values for the higher
twist terms ($A_2$=$-0.100 \pm 0.005$, $A_2$=$-0.0011 \pm 0.0008$, 
and $\chi^2/d.o.f.=1355$) if we use the standard MRS(R2) PDFs in the fit
(in this case, the nuclear corrections to the deuteron data are not applied).
However, the floating factors for the deuteron at high $x$ are higher than
the floating factors for the proton, as shown in Fig.~\ref{fig:floatNLO_nodu}.
This indicates that the $d$ quark distribution
in the standard MRS(R2) PDFs is indeed underestimated at high $x$.

\begin{table}[b]
\caption[]{The relative normalizations of the NMC and BCDMS data
to the SLAC data,
and major systematic error shift of the BCDMS data 
from the NLO and NNLO QCD analysis}
\label{ht-norm}
\begin{center}
\begin{tabular}{|c|c|c|c|}\hline
      &\hfil NMC(\%) p(d)   &  BCDMS(\%) p(d)  &  BCDMS $\lambda$ \\ \hline
NLO           &\hfil  1.0 ( 0.5 )  &  -4.5 ( -2.5 )    & 1.8 \\
NNLO          &\hfil  1.9 ( 1.3 )  &  -2.7 ( -0.8 )    & 1.2  \\ \hline
\end{tabular}
\end{center}
\end{table}

The magnitude of the higher twist terms extracted
in this NLO analysis
 is  almost half of the size from
a previous NLO analysis of SLAC/BCDMS
data~\cite{Virchaux},
because that analysis was based on \alfs(\mztwo) $=0.113$,
while \alfs(\mztwo) $=0.120$ in the MRS(R2) PDF,
which is close to the current world average.
Within the renormalon model, 
our fits  can also be used to estimate the size of the higher twist effects
in $xF_3$ [e.g. the Gross-Llewellyn Smith (GLS) sum rule] in NLO,
as the higher twist terms in
$F_2$ and $R$ are related to the higher twist terms in $xF_3$.
However, since the QCD predictions for the GLS sum rule have been
calculated to higher order on QCD, it is important to understand
if these significant higher twist contributions extracted in the
NLO analysis originate from real $1/Q^2$ and $1/Q^4$ terms,
or do they partly describe the missing higher order NNLO terms.

\begin{figure}
\centerline{\psfig{figure=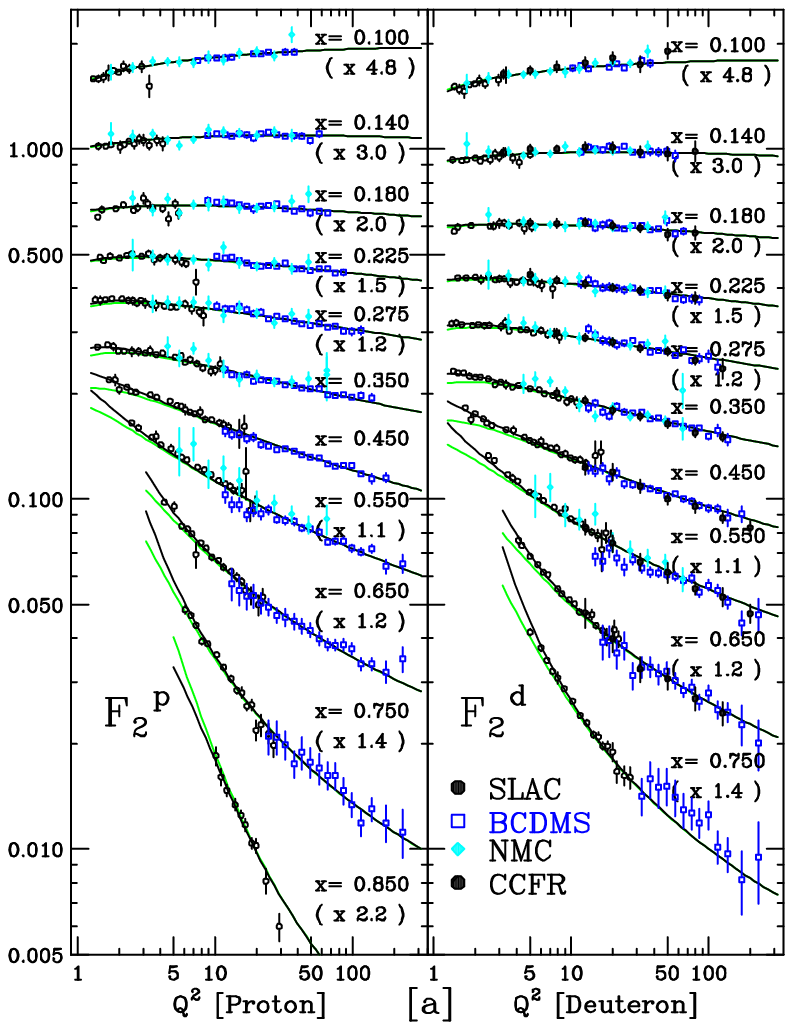,width=3.1in,height=4.1in}}
\centerline{\psfig{figure=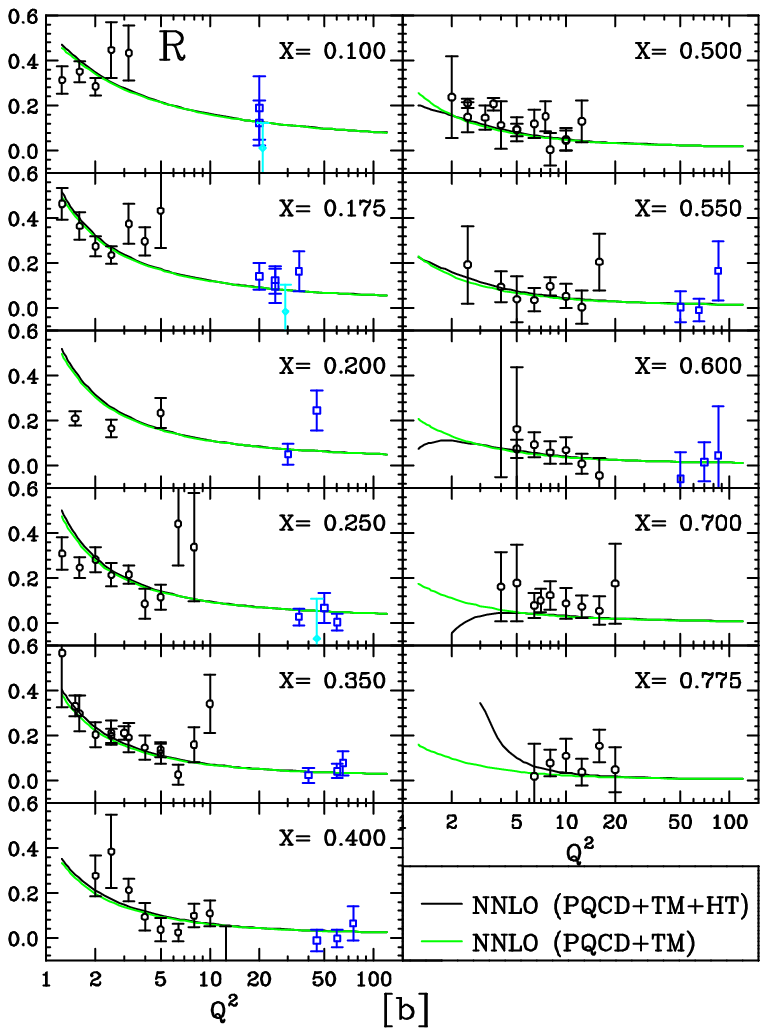,width=3.1in,height=4.1in}}
\caption{
Comparison of data with the QCD NNLO+TM+HT fit
(the HT renormalon model, $f^{NNLO}(x)$,
and the modified  MRS(R2) PDFs are used).
 The CCFR neutrino data are also shown  for comparison.
(a) Comparison of $F_2$ data and the NNLO+TM prediction
with and without HT contributions. 
(b) Comparison of $R$ data and the NNLO+TM prediction
with and without  HT contributions.}
\label{fig:disht2}
\end{figure}

\begin{figure}
\centerline{\psfig{figure=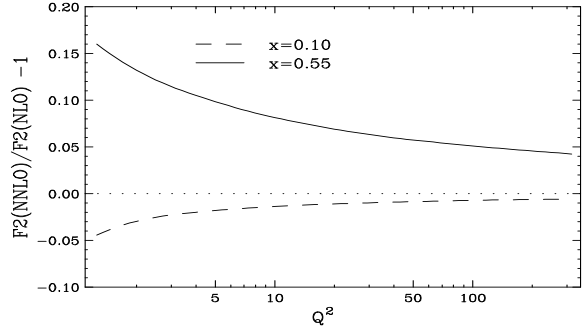,width=3.0in,height=1.7in}}
\caption{The $Q^2$ dependence of the NNLO contributions to $F_2$
for two representative values of $x$. The $Q^2$ dependence of the NNLO
contributions appears to be similar to that of the higher twist
contributions extracted in the NLO analysis.}
\label{fig:nnloterms}
\end{figure}

\begin{figure}
\centerline{\psfig{figure=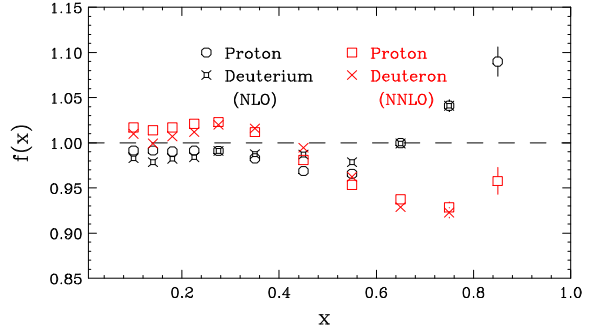,width=3.0in,height=1.7in}}
\caption{The floating factor $f^{NNLO}(x)$
 as a function of $x$ from the NNLO analysis. The $f^{NLO}(x)$ from the NLO 
analysis is also shown for comparison.
The ratio $f^{NNLO}(x)/f^{NLO}(x)$ corresponds
to the ratio of NNLO to NLO PDFs.}
\label{fig:floatNNLO}
\end{figure}

\section{NNLO QCD analysis}

We now proceed to repeat the analysis, but now include the NNLO
contributions to $F_2$ and $R$. 
Since the NNLO PDFs are not available,
the approach taken in our analysis
is that we input NLO PDFs into the NNLO expression for $F_2$ and $R$.
The NNLO theoretical predictions are compared to the data, and as
was done in our previous NLO fit, we extract
the higher twist coefficients $A_2$ and $A_4$ (from the $Q^2$ dependence
comparison) and the floating factor $f^{NNLO}(x)$ (from the $x$ dependence
comparison). 
The higher twist coefficients $A_2$ and $A_4$ can be interpreted
as representing both  the higher twist terms and the
difference in the $Q^2$ dependence between the input NLO PDFs
and the unknown NNLO PDFs. The ratio $f^{NNLO}(x)/f^{NLO}(x)$ 
can be interpreted as the ratio of NNLO to NLO PDFs as a function of $x$.

 Figure~\ref{fig:disht2} shows that the
fit including the NNLO contributions yields a good   
($\chi^2/d.o.f.=1375/926$, $\chi^2=1204(F_2)+171(R)$ )
description of the $x$ and $Q^2$ dependence for $F_2$ and $R$.
The relative normalizations of the NMC and BCDMS data
and major systematic error shift of the BCDMS data 
are also shown in Table~\ref{ht-norm}.
The extracted values of $A_2$ and $A_4$
are $-0.0065 \pm 0.0059$  and $-0.012 \pm 0.0008$, respectively.
The contribution of $A_2$ is found  to be negligible in this
NNLO analysis, and the $A_4$ term is small but finite.
These results indicate that 
most of the higher twist contributions extracted in the NLO fit
at low $Q^2$ appear to originate from the missing NNLO terms.
We also achieve the same conclusion even if we adopt the empirical
higher twist model~\cite{dupaper} by using only $F_2$ data.
In fact, Fig.~\ref{fig:nnloterms} shows that 
the $Q^2$ dependence of the NNLO contributions to $F_2$
is similar to that of the higher twist terms extracted 
in the NLO analysis. 
The small contribution of the higher twist terms to $F_2$ and $R$
in the NNLO analysis
also indicates that the higher twist contributions
to the GLS sum rule are very small.
The above values of $A_2$ and $A_4$ yield
a fractional contribution
to the pQCD GLS sum rule  of $-0.009/Q^2-0.023/Q^4$. Similar
conclusions for the GLS sum rule (from an global analysis of
data on $xF_3$ only) have been reported~\cite{russian} elsewhere.

Since the $Q^2$ dependence of the data is well described
by the fit, the results also imply that the $Q^2$ dependence
of the NNLO and NLO PDFs are mostly the same.  This conclusion
is in agreement with most recent estimates
of the three-loop splitting functions
for the NNLO PDFs by van Neerven and Vogt~\cite{nnlo_van}.
The ratio of NNLO to NLO PDFs as a function of $x$
can be obtained from the ratio $f^{NNLO}(x)/f^{NLO}(x)$.
Figure~\ref{fig:floatNNLO} shows the floating factor
$f^{NNLO}(x)$ as a function of $x$ for
the proton and deuteron respectively. At low $x$, 
the $f^{NNLO}(x)$ is few percent higher than $f^{NNO}(x)$, 
thus indicating that NNLO cross sections for the top quark, $W$ and $Z$
production will be somewhat increased if our NNLO corrections
to the NLO PDFs are used for the NNLO PDFs (instead of NLO PDFs).
For example, in the case of $Z$ production at the Tevatron,
the total theoretical cross section 
would be about $5\%$ higher (which will bring the theory into
closer agreement with the data~\cite{Zprod}).
At large $x$
the NNLO PDFs may be about $10\sim15\%$ lower than the NLO PDFs
mainly due to the two-loop coefficient functions.
The NNLO contributions to $R$ appear to account
for most of the higher twist effects extracted in the NLO fit.
Since the NNLO terms are important at
small $x$ (especially for $R$, in which the overall level
of $F_2$ in NNLO cancels out), we also conclude that
 with the increasing
precision of the data from HERA, these terms should no longer
be neglected.  Note that our conclusions are not sensitive
to the choice of model that is used to describe the behaviour
of the higher twist terms.
The same conclusions are obtained if we use an empirical
higher twist model~\cite{dupaper} to fit the $F_2$ data.
%We will be reporting on 
%our work~\cite{rpaper} on $R$ at low values of $x$ in a longer
%communication.

\section{Conclusion}

In conclusion
we find that a next-to-next-to-leading order (NNLO)
 analysis of $F_2$ and  $R$ shows  that
most of the higher twist contributions extracted in the NLO fit 
at low $Q^2$ appear to originate from the missing NNLO terms.
Within the renormalon model, the higher twist terms in
$F_2$ and $R$ are related to the higher twist terms in $xF_3$.
Therefore, the results imply that the higher twist contributions
to the GLS sum rule for $xF_3$ are very small. 
The analysis indicates that NNLO PDFs are a few percent higher than
NLO PDFs at small $x$ ($x$ near 0.1) which is the region that
contributes the most to top quark, $W$ and $Z$ production cross sections
at the Tevatron. At higher $x$ the NNLO PDFs are about $10\%$
lower than the NLO PDFs. 
The estimated ratio of NNLO to NLO
PDFs from our
analysis may be used to estimate the additional uncertainty
in NNLO calculations which originates from the fact that NNLO
PDFs are not currently available.
Our results 
are in agreement with most recent estimates
of the three-loop splitting functions
for the NNLO PDFs by van Neerven and Vogt~\cite{nnlo_van}.


\begin{thebibliography}{5}

%
\bibitem{MRSR2}
A.D. Martin {\em et al}.,\pl{B 387}{419}{1996} 
%
%\bibitem{CTEQ3M}
%H.L. Lai {\em et al}., \prev{D 51}{4723}{1995}
\bibitem{CTEQ4M}
H.L. Lai {\em et al}., \prev{D 55}{1280}{1997}
%
\bibitem{GRV94}
M. Gluck {\em et al.}, {\it Zeit. Phys.} {\bf C67}, 433 (1995).

\bibitem{NLO}
%G. Altarelli and G. Martinelli, Phys. Lett. {\bf 76B}, 89 (1978).
G. Altarelli and G. Parisi, Nucl. Phys. {\bf B 126}, 298 (1977);
V. N. Gribov and L. N. Lipatov, Sov. J. Nucl. Phys. {\bf 15}, 438 (1972);
Yu L. Dokshitzer Sov. Phys. JETP {\bf 46}, 641 (1977).
% The above is reference to DGLAP
%
\bibitem{dupaper}
U. K. Yang and A. Bodek, Phys. Rev. Lett. {\bf 82}, 2467 (1999).
%
\bibitem{GPtm}
H. Georgi and H.D. Politzer, {Phys. Rev.} {\bf D 14}, 1829 (1976).
%
\bibitem{renormalon}
M. Dasgupta and B.R. Webber, \pl{B 382}{273}{1996} 

\bibitem{NNLO}
 E. B. Zijlstra and W. L. van Neerven, Nucl. Phys.
 {\bf B 383}, 525 (1992); Phys. Lett. {\bf B 273}, 476 (1991); 
{\bf B 272}, 127 (1991); 
J. Sanchez Guillen {\em et al}., Nucl. Phys. {\bf B 353}, 337 (1991);
L. H. Orr and W. J. Stirling, Phys. Rev. Lett. {\bf 66}, 1673 (1991).
%

%
\bibitem{SLACF2}
L.W. Whitlow  {\em et al}., \pl{B 282}{475}{1992} 
%
\bibitem{BCDMSF2}
A.C. Benvenuti {\em et al}., Phys. Lett. {\bf B 223}, 485 (1989);
A.C. Benvenuti {\em et al}., \pl{B 237}{592}{1990} 
%
\bibitem{NMCF2}
M. Arneodo {\em et al}., \np{B 483}{3}{1997}
% The above is correct checked NMC D and H reference

\bibitem{SLACres}
P. Bosted  {\em et al}., \prev{D 49}{3091}{1994}
%SLAC high x data

\bibitem{GOMEZ}
J. Gomez {\em et al}., \prev{D 49}{4348}{1994} 

%
%
\bibitem{CCFRF2}
W.G. Seligman {\em et al}., \prl{79}{1213}{1997}
% CCFR data
%
%
\bibitem{Virchaux}
M. Virchaux and A. Milsztajn, \pl{B 274}{221}{1992} 
%Old SLAC/BCDMS higher twist analysis


% Russian XF3 analysis
\bibitem{russian}
S. I. Alekhin and A. L. Kataev,  Phys.Lett. {\bf B452}, 402 (1999).


\bibitem{nnlo_van}
W.L. van Neerven and A. Vogt, INLO-PUB 14/99, hep-ph/9907472
(July 1999).

% CDF Z production
\bibitem{Zprod}
T. Affolder {\em et al}., FERMILAB-PUB-99-220-E, 
submitted to Phys. Rev. Lett.

\end{thebibliography}
\end{document}